\documentclass[
reprint,
pra,
]{revtex4-1}

\usepackage{makeidx}
\usepackage{amsmath,amssymb,amsfonts}
\usepackage{natbib}
\usepackage{braket}
\usepackage{graphicx}
\usepackage{dcolumn}
\usepackage{bm}
\usepackage[
]{geometry}

\begin{document}


\title{Comments on the Fractal Energy Spectrum of Honeycomb Lattice with Defects}

\author{Yoshiyuki Matsuki}
\email{ymatsuki@het.phys.sci.osaka-u.ac.jp}
\author{Kazuki Ikeda}%
\email{kikeda@het.phys.sci.osaka-u.ac.jp}
\affiliation{ 
Department of Physics, Osaka University, Toyonaka, Osaka 560-0043, Japan
}%

\date{\today}

\begin{abstract}
We address the energy spectrum of honeycomb lattice with various defects or impurities under a perpendicular magnetic field. We use a tight-binding Hamiltonian including interactions with the nearest neighbors and investigate its energy structure for two different choices of point defects or impurities. In the first case, we fix a unit cell consisting of 8 lattice points and survey the energy eigenvalues in the presence of up to 2 point defects. Then it turns out that the existence of the fractal energy structure, called Hofstadter's butterfly, depends on the choice of defect pairs. In the second case, we extend the size of a unit cell which contains a single point defect and up to 32 lattice points. The fractal structures indeed appear for those cases and there exist a robust gapless point in the $E=0$ eV line without depending on both the size of unit cells and the shape of lattices. Therefore we keep an immortal butterfly since such a robust point corresponds to the center of a butterfly. Consequently we predict that the presence of a butterfly in a graph is equivalent to that of fractality.       
\end{abstract}
\maketitle
\section{Introduction} 
It has been widely known that the graph of the spectrum over a wide range of rational magnetic fields passing through a two-dimensional lattice cell shows fractal, called the Hofstadter butterfly \cite{PhysRevB.14.2239}, which is confirmed by various experimental studies \cite{cite-key1,cite-key2,Hunt1427}. This intriguing behavior of Bloch electrons have long attracted the major interest of researchers from various perspectives. In a two-dimensional system, the Hofstadter problem is addressed for various choice of lattices: general Bravais lattice \cite{2017PhRvA..95f3628Y}, buckled graphene like materials \cite{PhysRevB.91.235447} and square lattice with next-nearest-neighbor hopping \cite{PhysRevB.42.8282}. More recently Hofstadter's butterfly appears in relation to the quantum geometry \cite{Hatsuda:2016mdw,Hatsuda:2017zwn} and it is also surveyed from a perspective of mathematical physics \cite{Ikeda:2017ztr,Ikeda:2017uce}. This problem is extended to higher dimensional cases \cite{PhysRevLett.86.1062,PhysRevB.69.033202,ptu144} and it is known that analog of Hofstadter’s butterfly exists in more general systems. 

All of the above theoretical studies are based on the perfect lattice structures, therefore a simple question "Does the fractal energy structure appear even in systems with defects?" comes to our mind. As well as investigating the origin of the fractal nature, we aim at giving a positive answers to this question. Apparently it is a formidable task to find such a fractal energy spectrum in a system with defects since they usually make energy bands gapped. Indeed, as we report in this article the energy spectrum highly depends on a choice of defect pairs and analogue of the Hofstadter butterfly exists only under appropriate conditions. It is briefly reported that the honeycomb lattice with a single point defect in a unit cell accommodates the fractal spectrum with large band gaps nearby the $E=0$ eV line (Fig.\ref{E}) and point defects are responsible for modifying the Hofstadter butterfly structure \cite{PhysRevB.85.235414}. As a succeeding study, we investigate more on roles of defects or impurities in order to enhance our knowledge on its fractal natures. Though it is obvious that the fractal structure cannot be obtained if too many defects are introduced into a unit cell, as a non-trivial result, we find that the large band gaps reported in the previous paper \cite{PhysRevB.85.235414} are closed and the graph of the energy spectrum looks like symmetric rather than fractal when another atom is vacant so that the honeycomb lattice accommodates line defects (Fig.\ref{2}). In addition, even for several enlarged unit cells, we also observe such a robust gapless point in the $E=0$ eV level that exists without depending on a choice of the unit cell size. Moreover, it can be found at the center of the butterfly shape ($\phi=0.5$ in Fig.\ref{butt}) and it is crucial to the fractal nature of the graph.  

This article is orchestrated as follows. In Section \ref{i} we explain our problem setting. We use the tight-binding Hamiltonian including the nearest neighbor interactions of electrons in enlarged unit cells with point defects or impurities. In Section \ref{ii}, we present a number of graphs obtained by numerical calculation for various choice of defects and for several enlarged unit cells. We also give several theoretical reasons for the obtained family of energy structures and consider the origin of the fractal nature from a purely linear algebraic perspective. This article is concluded in Section \ref{iii} with some proposed future works.

\section{\label{i}Model and Formulation} 
We introduce the magnetic field into the 2 dimensional lattice system by the Peierls substitution\cite{Peierls1933}
\begin{align}
\begin{aligned}
\ket{\bf{R}_i}\bra{\bf{R}_j} \rightarrow \mathrm{e}^{i\Theta_{ji}^{mag}}\ket{\bf{R}_i}\bra{\bf{R}_j},
\end{aligned}
\end{align}
where $\Theta_{ji}^{mag}=-\frac{e}{\hbar}\int_{\bf{R}_j}^{\bf{R}_i} \bf{A} \cdot d\bf{l}$ In the Landau gauge $(0,Bx,0)$, it corresponds to 
\begin{eqnarray}
\Theta_{ji}^{mag}= -\pi \phi({\bf{R}_j}+{\bf{R}_i})\cdot \hat{x}\left(({\bf{R}_i}-{\bf{R}_j})\cdot \hat{y}\right),
\end{eqnarray}
where ${\bf{R}_i}$ is a position vector specifying the atom, ${\bf{R}_i}=m_{\alpha}{\bf{a_{1}}}+n_{\alpha}{\bf{a_{2}}}$ with labels  $\alpha$ of the atoms in the enlarged unit cell, $\phi=BS/\phi_{0}$, $\phi_{0}=h/e$ and $S$ is the area of the enlarged unit cell. \\

\begin{figure}[htbp]
\centering
\includegraphics[width=6cm, bb=0 0 360 361]{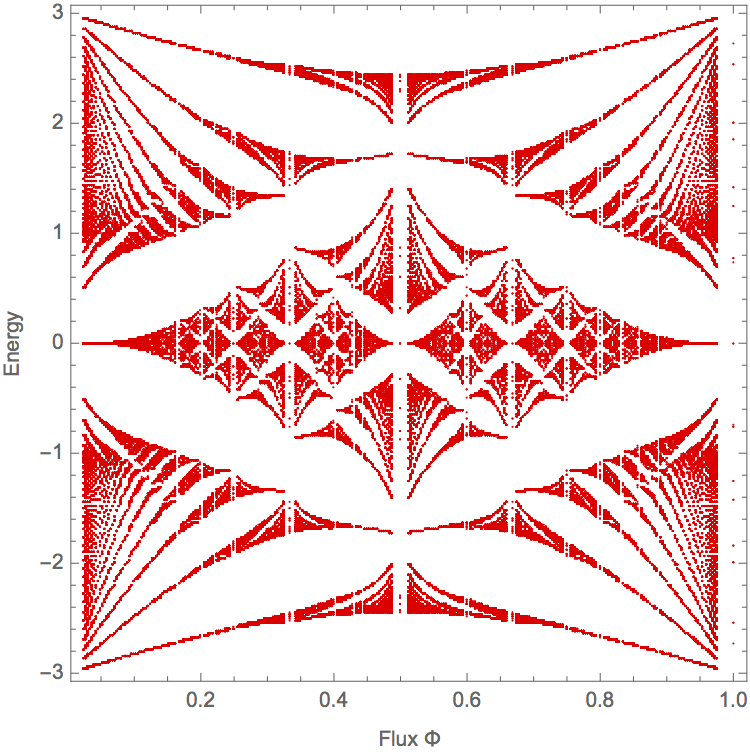}
\caption{\label{butt}Energy spectrum plotted over a wide range of $\phi$ on a pure honeycomb lattice.}
\end{figure}

\begin{figure*}[htbp]
\begin{minipage}{0.3\hsize}
\begin{center}
\includegraphics[width=4cm, bb=0 0 584 593]{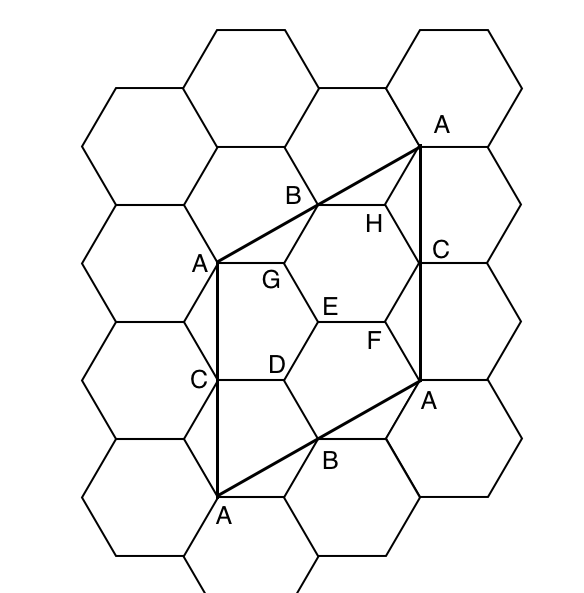}
\end{center}
(a)
\label{fig:one}
\end{minipage}
\begin{minipage}{0.3\hsize}
\begin{center}
\includegraphics[width=4cm, bb=0 0 430 474]{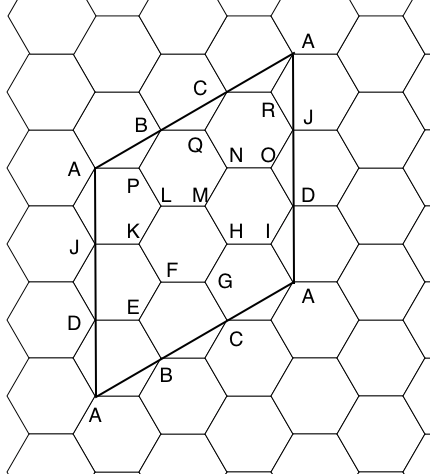}
\end{center}
(b)
\label{fig:two}
\end{minipage}
\begin{minipage}{0.3\hsize}
\begin{center}
\includegraphics[width=4cm, bb=0 0 430 474]{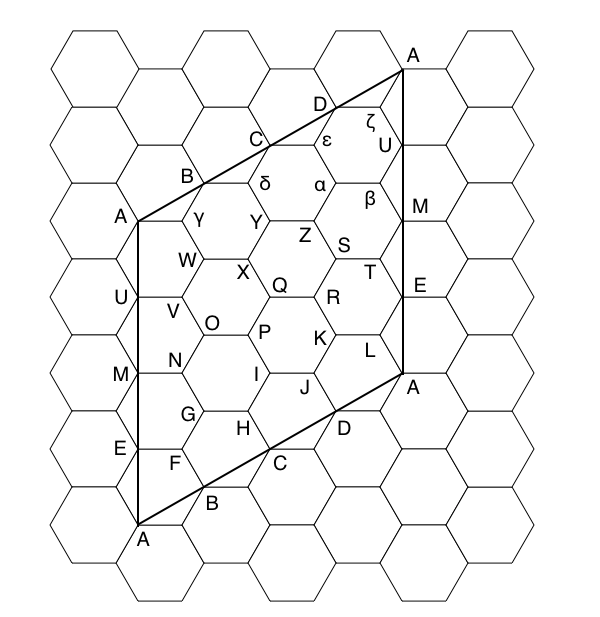}
\end{center}
(c)
\label{fig:three}
\end{minipage}
\caption{\label{cell}Enlarged unit cells of the honeycomb lattice. Each of them consists of 8 (a), 18 (b) and 32 (c) atoms respectively. For example, the lattice vector are ${\bf{a_{1}}}=3a\hat{x}+\sqrt{3}a\hat{y},{\bf{a_{2}}}=2\sqrt{3}a\hat{y}$ in the case of (a).} 
\end{figure*}

Under this setup, our system can be written by means of a new matrix equation which can be called the generalized Harper equation\cite{0370-1298-68-10-305}
\begin{eqnarray}
E{\bf{\Psi_{m}}}=U_{m}{\bf{\Psi_{m}}}+V_{m}{\bf{\Psi_{m+1}}}+W_{m}{\bf{\Psi_{m-1}}}. 
\end{eqnarray}
Using matrix representation, we rewrite it as
\begin{widetext}
\begin{eqnarray}\label{H}
E\left[ 
\begin{array}{r}
{\bf{\Psi_{1}}}\\
{\bf{\Psi_{2}}}\\
\vdots\\
{\bf{\Psi_{q-1}}}\\
{\bf{\Psi_{q}}}
\end{array}
\right]=
\left(
\begin{array}{rrrrrrr}
U_{1} & V_{1} & 0 & 0 & \ldots & 0& W_{1}^{\ast} \\
W_{2} & U_{2} & V_{2} & 0 & 0 &\ldots & 0 \\
0 & W_{3} & U_{3} & V_{3} & 0 & \ldots & 0  \\
\vdots & \vdots & \ddots & \ddots & \ddots & \ldots & \vdots \\
V_{q}^{\ast} & 0 & 0 & \ldots & 0 & W_{q} & U_{q}  
\end{array}
\right)
\left[ 
\begin{array}{r}
{\bf{\Psi_{1}}}\\
{\bf{\Psi_{2}}}\\
\vdots\\
{\bf{\Psi_{q-1}}}\\
{\bf{\Psi_{q}}}
\end{array}
\right],
\end{eqnarray}
\end{widetext}
where $U_m,V_m,W_m$ are certain matrixes and $W_{1}^{\ast}=W_{1}\mathrm{e}^{-ik_{x}a_{1x}q}$,~$V_{q}^{\ast}=V_{q}\mathrm{e}^{-ik_{x}a_{1x}q}$ $\big(-\frac{\pi}{q}$ $\leq$ $k_{x}$ $\leq$ $\frac{\pi}{q}$ ,$-\pi$ $\leq$ $k_{y}$ $\leq$ $\pi\big)$. Note that the index $m$ is periodic in $q$ with a period $q$ therefore wave functions are written
\begin{eqnarray}
\psi_{m+q}=\mathrm{e}^{ik_{x}a_{1x}q} \psi_{m}
\end{eqnarray}
by Bloch's theorem. For example, if the unit cell (a) in Fig.\ref{cell} is preferred, then $\bf{\Psi_{m}}$ has the form 
\begin{eqnarray}
{\bf{\Psi_{m}}}=\left[
\begin{array}{r}
\psi^{A}_{m}\\
\psi^{B}_{m}\\
\psi^{C}_{m}\\
\psi^{D}_{m}\\
\psi^{E}_{m}\\
\psi^{F}_{m}\\
\psi^{G}_{m}\\
\psi^{H}_{m}\\
\end{array}
\right]
\end{eqnarray}
and $U_{m}, V_{m}, W_{m}$ are $8\times 8$ matrices. Then solving the characteristic equation $\det(E-H)=0$ with respect to each $\phi$, where $H$ is defined by the left hand side of the equation \eqref{H}, we obtain the butterfly shape (Fig.\ref{butt}). 

We investigate energy spectra of Bloch electrons in the honeycomb lattice whose unit cell is chosen in three different ways as displayed in Fig.\ref{cell}, and consider its dependence on both position- and density-dependence of impurities or defects. In the first study, we treat the atom at $E$ as a defect and introduce additional impurities at $C$, $F$ and $H$, respectively. (Those three pairs are enough for our purpose since the honeycomb lattice has the $\pi/6$ rotation symmetry as well as the transition symmetry.) $(E,F)$ is a nearest neighbor pair of defects, $(E, C)$ is a next-nearest neighbor pair of defects, and $(E, H)$ is a third nearest neighbor pair of defects. In the second study, we extend the size of unit cells (from (a) to (c) in Fig.\ref{cell}) and consider the size dependence of the energy spectrum. Each unit cell contains a single defect ($E$ in (a), $L$ in (b) and $Q$ in (c)) and purity of the honeycomb lattice would get close to a realistic situation (0.125 (a), 0.056 (b), 0.031 (c)). Throughout this article, all interactions among Bloch electrons are treated as the nearest neighbor interactions. 

\section{\label{ii} Results and Discussions}
\subsection{2 Defects in a Unit Cell} 
\begin{figure}[h!]
\centering
\includegraphics[width=6cm, bb=0 0 360 361]{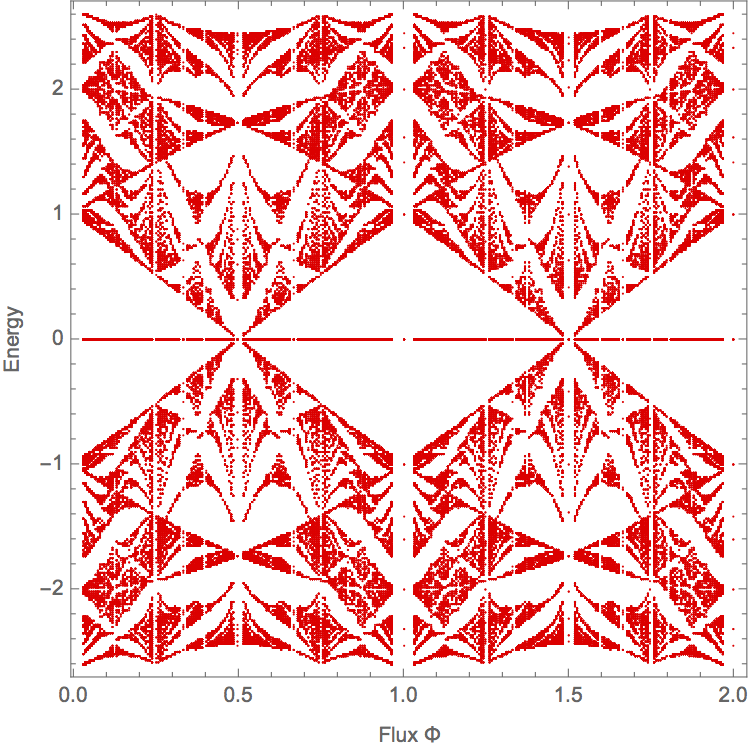}
\caption{\label{E}Spectrum for a honeycomb lattice whose unit cell has a single defect.}
\end{figure}

We first begin with the case where a single defect exists at $E$ in a unit cell (Fig. \ref{E}). Compared with studies on Hofstadter's butterfly on a perfect lattice, a similar fractal nature can be found. Note that there are gapless points at $\phi=k+1/2$ ($k\in\mathbb{N}_{\ge 0}$) in the $E=0$ eV line and large band gaps having triangular shapes whose boundaries are formed by $\cos$ curves are formed elsewhere.  

\begin{figure*}
\begin{minipage}{0.31\hsize}
\begin{center}
\includegraphics[width=4.5cm, bb=0 0 360 361]{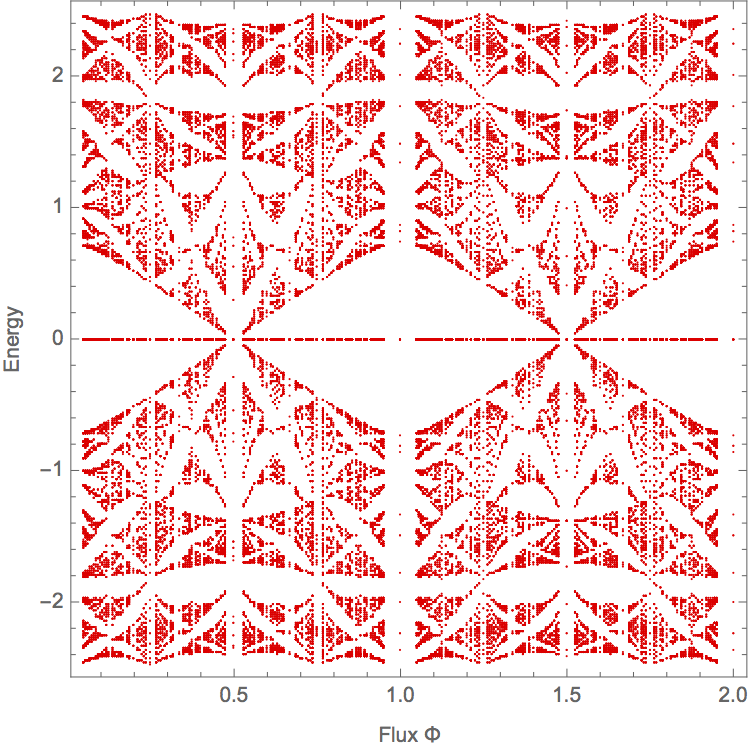}
\end{center}
$H^F_{pp\pi}=2/3H_{pp\pi}$
\label{fig:one}
\end{minipage}\vspace{1mm}
\begin{minipage}{0.31\hsize}
\begin{center}
\includegraphics[width=4.5cm, bb=0 0 360 361]{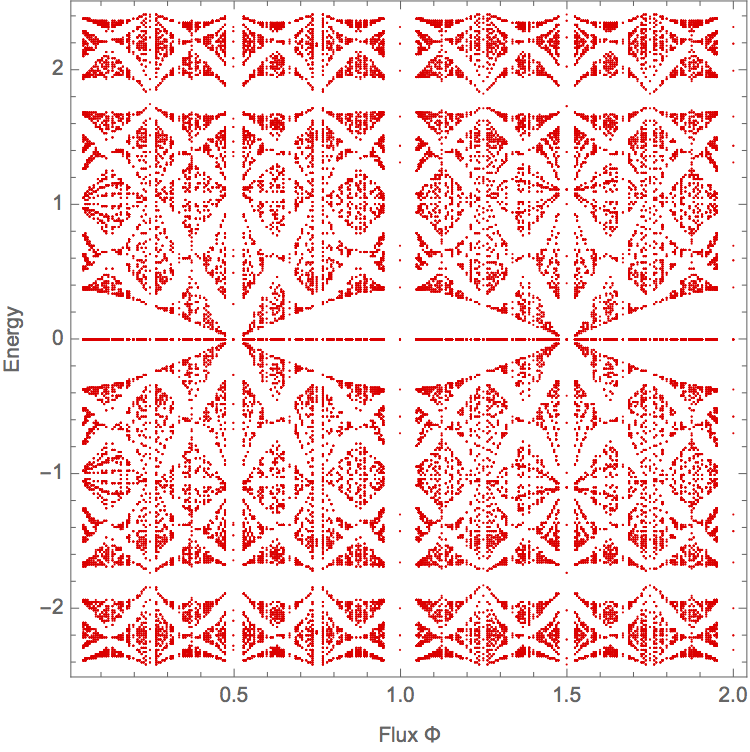}
\end{center}
$H^F_{pp\pi}=1/3H_{pp\pi}$
\label{fig:two}
\end{minipage}\vspace{1mm}
\begin{minipage}{0.31\hsize}
\begin{center}
\includegraphics[width=4.5cm, bb=0 0 360 361]{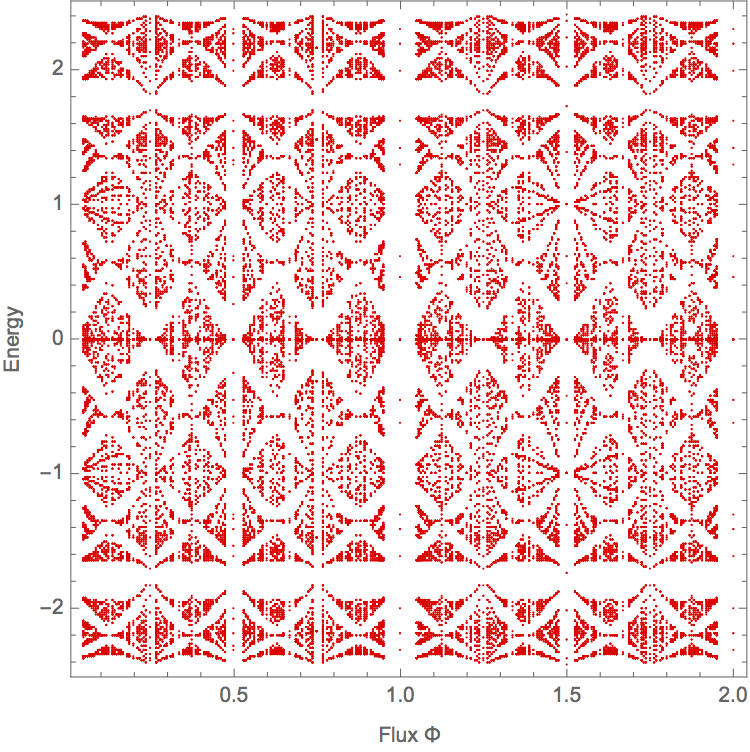}
\end{center}
$H^F_{pp\pi}=0$
\label{fig:three}
\end{minipage}\vspace{1mm}
\begin{minipage}{0.31\hsize}
\begin{center}
\includegraphics[width=4.5cm, bb=0 0 360 361]{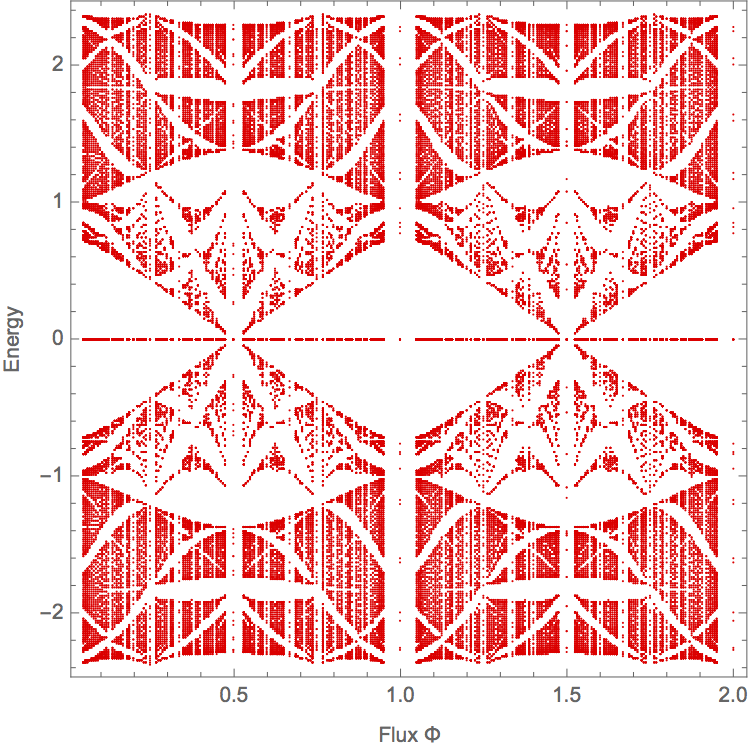}
\end{center}
$H^C_{pp\pi}=2/3H_{pp\pi}$
\label{fig:three}
\end{minipage}\vspace{1mm}
\begin{minipage}{0.31\hsize}
\begin{center}
\includegraphics[width=4.5cm, bb=0 0 360 361]{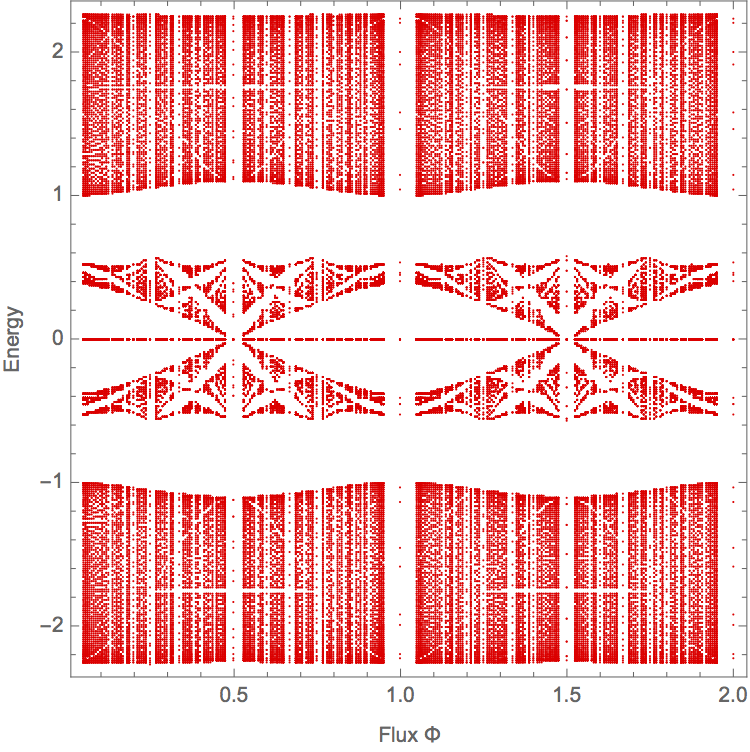}
\end{center}
$H^C_{pp\pi}=1/3H_{pp\pi}$
\label{fig:three}
\end{minipage}\vspace{1mm}
\begin{minipage}{0.31\hsize}
\begin{center}
\includegraphics[width=4.5cm, bb=0 0 360 361]{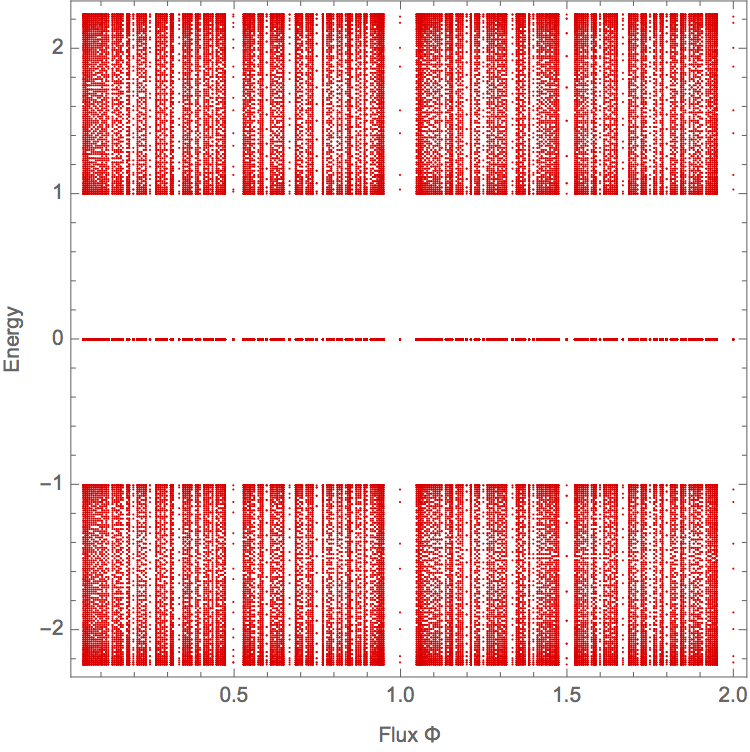}
\end{center}
$H^C_{pp\pi}=0$
\label{fig:three}
\end{minipage}\vspace{1mm}
\begin{minipage}{0.31\hsize}
\begin{center}
\includegraphics[width=4.5cm, bb=0 0 360 361]{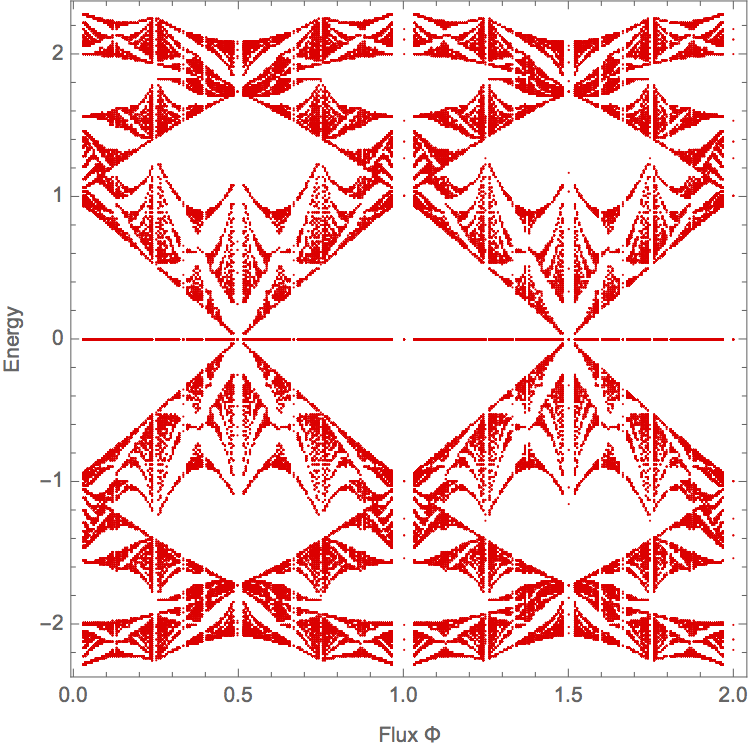}
\end{center}
$H^H_{pp\pi}=2/3H_{pp\pi}$
\label{fig:three}
\end{minipage}\vspace{1mm}
\begin{minipage}{0.31\hsize}
\begin{center}
\includegraphics[width=4.5cm, bb=0 0 360 361]{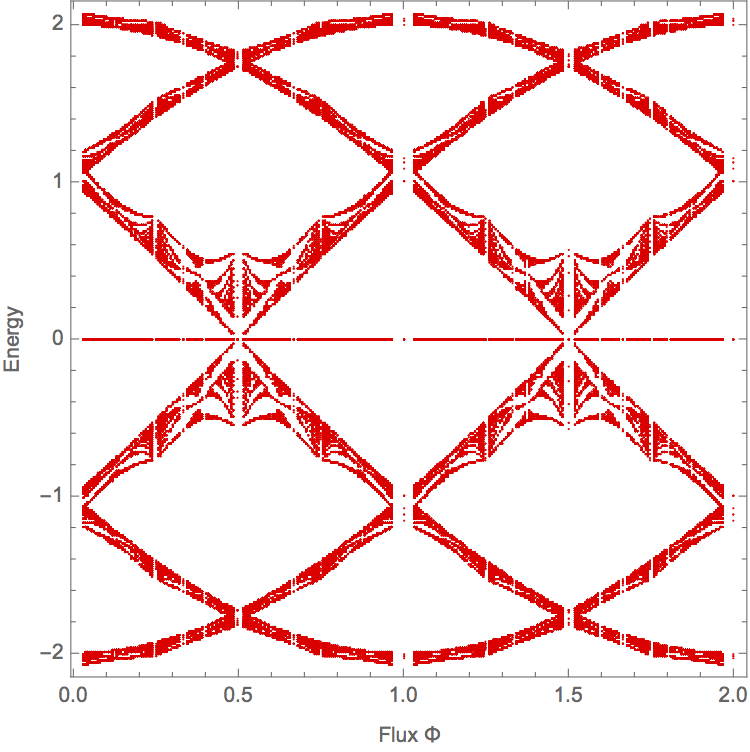}
\end{center}
$H^H_{pp\pi}=1/3H_{pp\pi}$
\label{fig:three}
\end{minipage}\vspace{1mm}
\begin{minipage}{0.31\hsize}
\begin{center}
\includegraphics[width=4.5cm, bb=0 0 360 361]{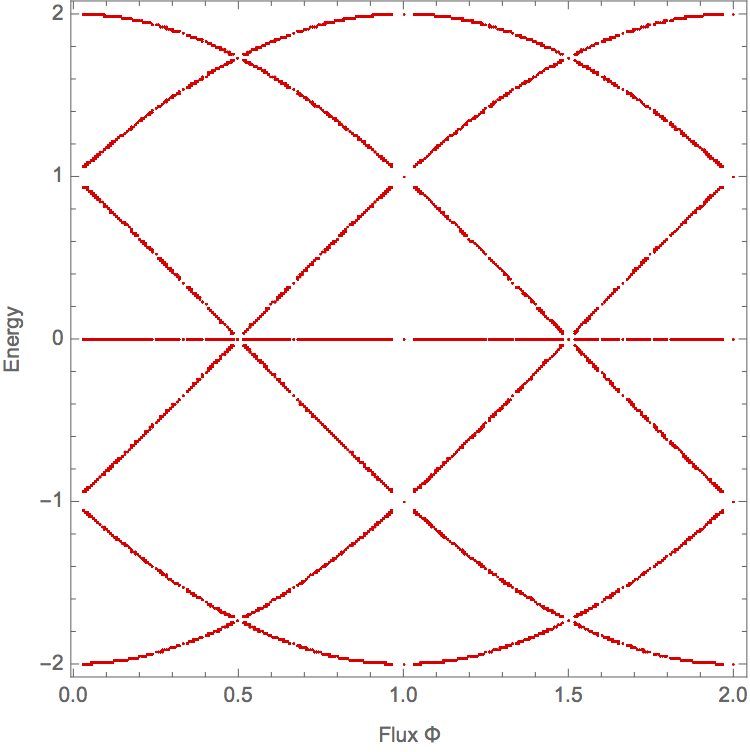}
\end{center}
$H^H_{pp\pi}=0$
\label{fig:three}
\end{minipage}\vspace{1mm}
\caption{\label{2}Graphs of energy spectrum over $\phi=4P/Q$, where $P$ and $Q$ are positive integers. Defects or impurities are introduced at points $(E,F)$, $(E, C)$ and $(E,H)$. The hopping parameter $H_{pp\pi}$ stands for the interaction of the $\pi$ orbital with the neighboring $\pi$ orbital. $H^E_{pp\pi}$ is set to 0 for all cases, i.e. $E$ is vacant.} 
\end{figure*}
\begin{figure}
\centering
\includegraphics[width=6cm, bb=0 0 360 361]{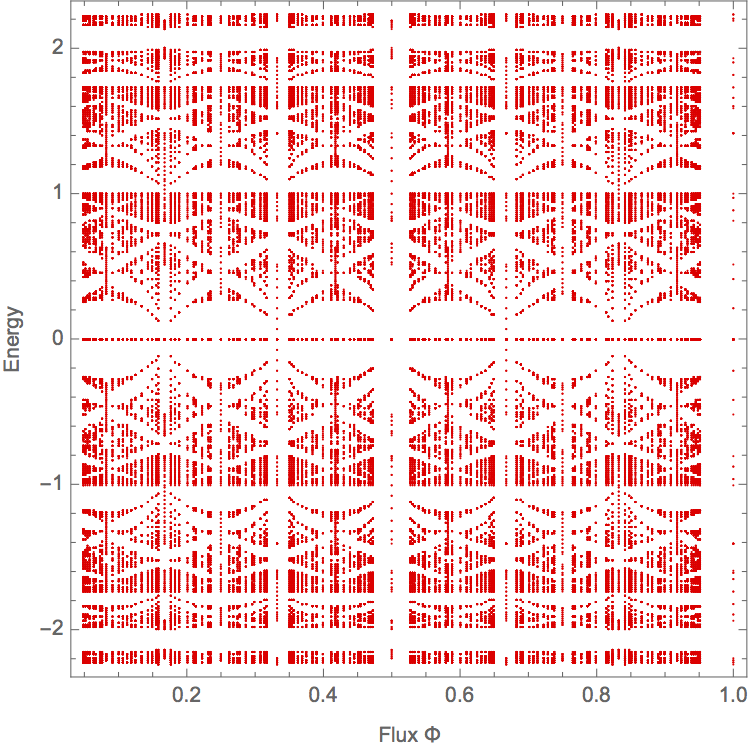}
\caption{\label{arm}Butterflies on the honeycomb lattice with armchair boundaries. We use the lattice (b) in Fig.\ref{cell} cut along $A,E,F,M,N,R$.}
\end{figure}

We next leave the atom at $E$ vacant and introduce additional impurities or defects. Graphs of the spectrum over a wide range of rational magnetic fields $\phi$ are displayed in Fig.\ref{2}. We denote by $H^F_{pp\pi}$ the interaction of the $\pi$ orbital of the atom at $F$ with the neighboring $\pi$ orbital. We first consider the case where an impurity is inserted at $F$.  According to the figures in Fig.\ref{2}, there are three different patterns based on positions of impurities. As interactions with the atom $E$ get weaken ($H^F_{pp\pi}\to0$), energy levels split and form gaps around $|E|\sim1.8~\text{eV}$. Consequently, energy levels within $\pm 1.8~\text{eV}$ ranges are modified and the blank area having a rugby ball shaped energy gap within $|E|<1.0~\text{eV}$ shrinks. If atoms at $E$ and $F$ are missed, then the graph acquires a new transition symmetry whose period is one-quarter of the original. When only a single impurity exists in a unit cell, those central band gaps are robust: it does not disappear for arbitrary non-zero value of hopping parameter $H^E_{pp\pi}$ \cite{PhysRevB.85.235414}. However this is not true if additional impurity is considered. In contrast to this case, flow of spectra formed in the process of $H^C_{pp\pi}\to0$ show completely different behavior. As seen in the figures of $H^C_{pp\pi}$, the arcwise sets of spectra in Fig.\ref{E} becomes boundaries which divide the chunks of spectra into three parts. As a result, all spectra within the range $|E|<0.8~\text{eV}$ in the figure of $H^C_{pp\pi}=1/3H_{pp\pi}$ approach to $E=0~\text{eV}$, namely the corresponding states get degenerated, and the graph of $H^C_{pp\pi}=0$ does not accommodate any fractal nature. Note that when atoms at $E$ and $C$ are missed, then the honeycomb lattice is no more arcwise connected and the boundaries are zigzag. One may naively think that this is consistent with the fact that Hofstadter's butterfly never appears on any one-dimensional lattice system, but this is not correct here. To confirm this, we also tried a case with armchair boundaries and the obtained graph somehow looks fractal (Fig.\ref{arm}).

Though there is a wide band gap in $|E|\le1~\text{eV}$, the other regions are gapless. This is in contrast to the case $H^F_{pp\pi}=0$, which has band gaps around $|E|=1.8~\text{eV}$. The last case $H^H_{pp\pi}\to0$ is also different from the above two cases. Energy bands localize around six independent stripes (or $\cos$ curves) which remain at $H^H_{pp\pi}=0$. They would correspond to energy bands of the six atoms living in the unit cell. It should be noted that there are band crossing points in the $E=0$ eV line throughout the process, and the fractal nature of graphs also observed for all non-zero $H^H_{pp\pi}$. According to the figures, it is only $H^H_{pp\pi}=0$ that the fractality becomes invisible. We may state things more formally as follows. Note that any square matrix $M$ over $\mathbb{C}$ is similar to a Jordan matrix (uniquely determined by $M$)
\begin{equation}
J=
\begin{pmatrix}
J_1&~&~&~\\
~&J_2&~&~\\
~&~&\ddots&~\\
~&~&~&J_k
\end{pmatrix},
\end{equation} 
where $J_i=J(\lambda_i, d_i)$ is labeled by its eigenvalue $\lambda_i$ and dimensions $d_i$, which are equal to the number of the corresponding degenerated quantum states. Thereby the smaller the number of Jordan cells, the more complicated the fractal structure becomes. The figure of $H^H_{pp\pi}=0$  would suggest that the minimum requirement for a graph to acquire fractality is 
\begin{equation}
k\ge \text{the number of atoms in a unit cell}.
\end{equation} 
In addition, it turns out by numerical calculation that the $H^C_{pp\pi}=0$ case is correspond to the case $\text{rank}J=\frac{1}{2}\text{dim}J$. Therefore, $\text{rank}J>\frac{1}{2}\text{dim}J$ will be also necessary to generate a fractal graph.

\subsection{Dependence on the Size of Unit Cells}
\begin{figure*}
\begin{minipage}{0.45\hsize}
\begin{center}
\includegraphics[width=6.5cm, bb=0 0 360 361]{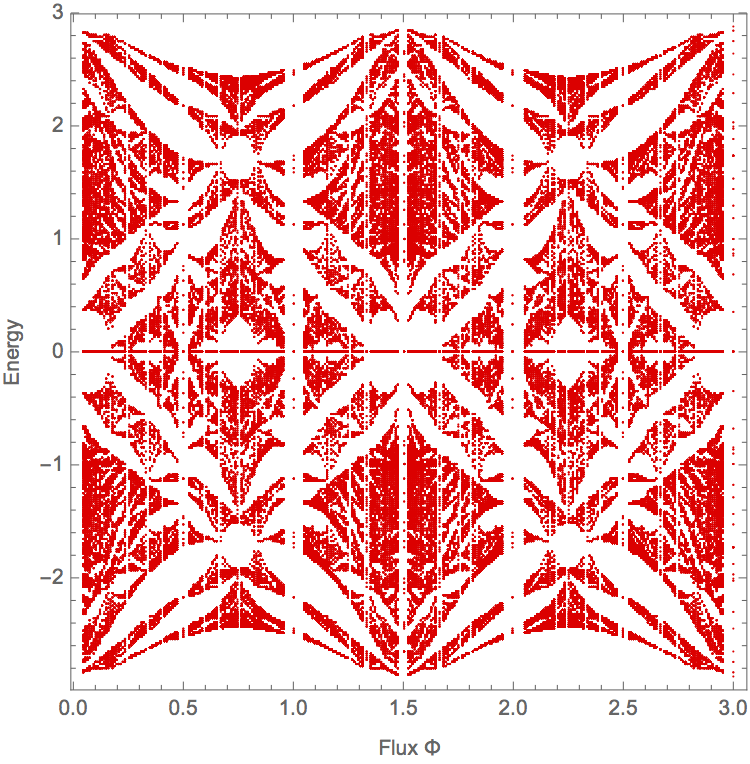}
\end{center}
(b-1)
\label{fig:three}
\end{minipage}\vspace{1mm}
\begin{minipage}{0.45\hsize}
\begin{center}
\includegraphics[width=6cm, bb=0 0 360 361]{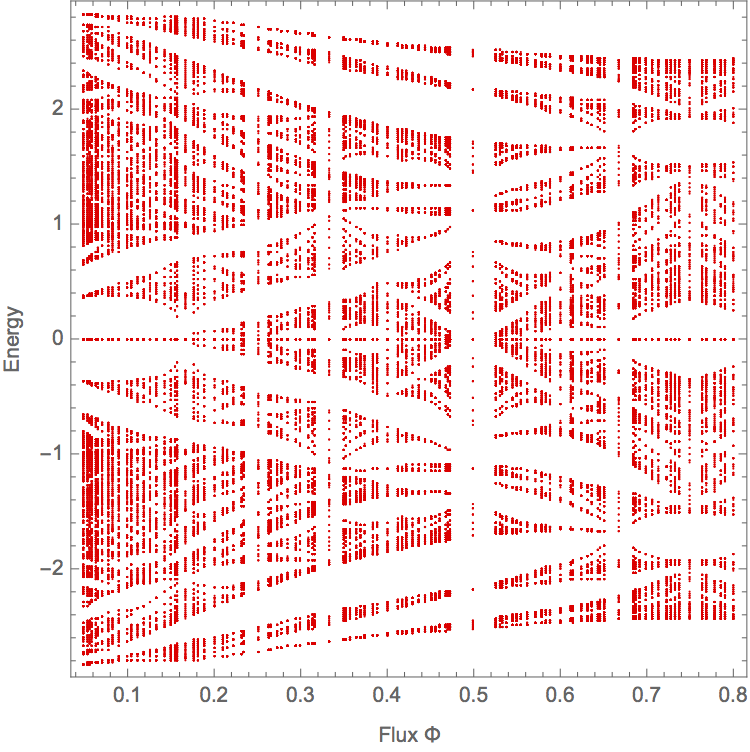}
\end{center}
(b-2)
\label{fig:three}
\end{minipage}\vspace{1mm}
\begin{minipage}{0.45\hsize}
\begin{center}
\includegraphics[width=6.5cm, bb=0 0 360 361]{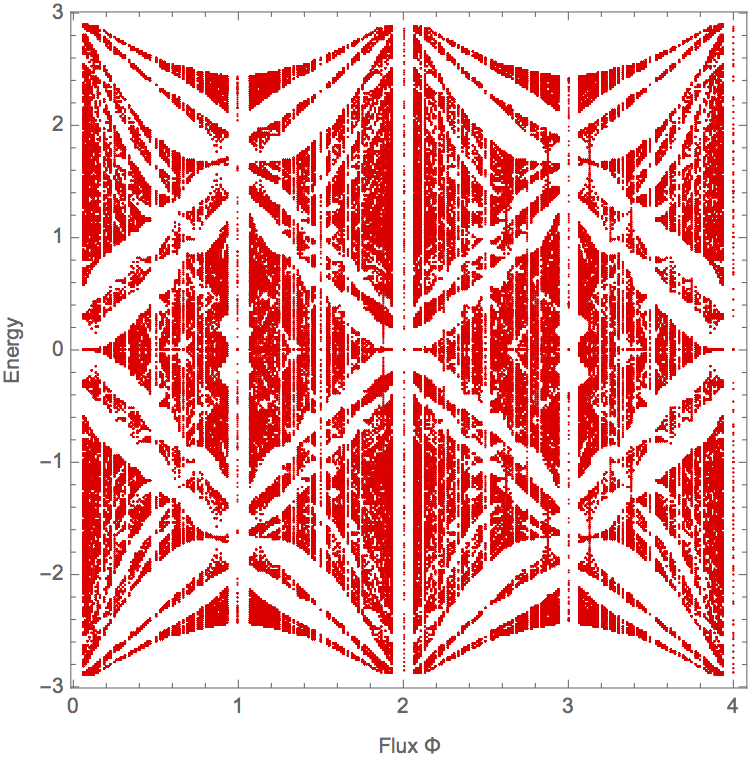}
\end{center}
(c-1)
\label{fig:three}
\end{minipage}\vspace{1mm}
\begin{minipage}{0.45\hsize}
\begin{center}
\includegraphics[width=6.5cm, bb=0 0 360 361]{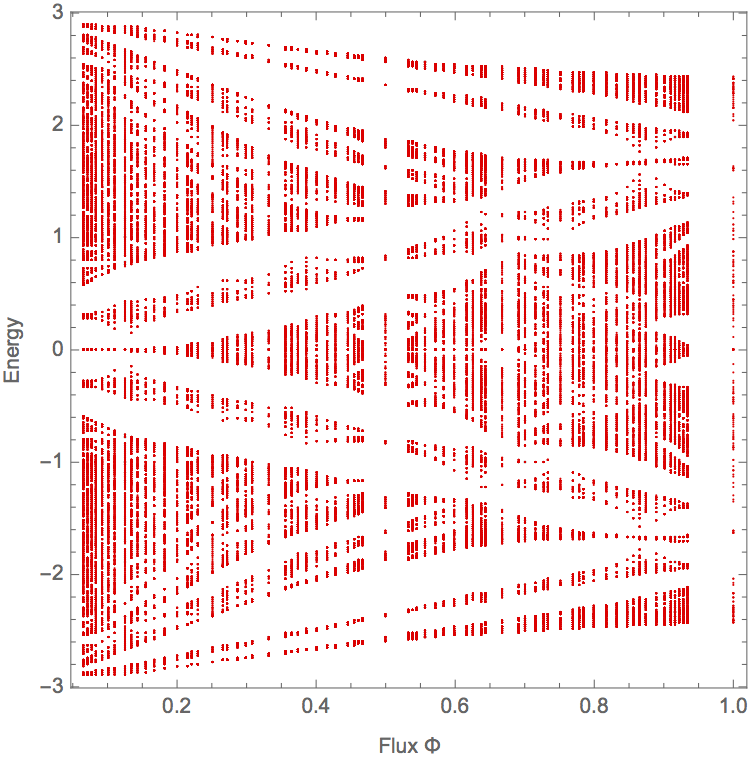}
\end{center}
(c-2)
\label{fig:three}
\end{minipage}\vspace{1mm}
\caption{\label{S} Spectra on the different enlarged unit cells. (b-1) corresponds to a two-period of the energy spectra considered on the unit cell (b) in Fig.\ref{cell} and the area around $\phi=0.5$ is picked up and displayed in (b-2). (c-1) and (c-2) are labeled in a same manner.} 
\end{figure*}

We next consider the unit cell size dependence of the energy spectrum. To address this problem, we extend the previous unit cell which contains 8 atoms as shown in Fig. \ref{cell}. The graphs of spectrum on unit lattice (b) in Fig.\ref{cell} with a single defect correspond to (b-1) and (b-2) in Fig.\ref{S}, and similarly defined for spectra on unit cell (c). Fractal structures can be found in either case. One significant aspect is that there are robust gapless points ($\phi=0.5$ for example) in the $E=0$ eV line even if a single defect is inserted in a unit cell, without depending on its size (see (b-2) and (c-2) for detail). Those points exist in the Hofstadter problem on the perfect honeycomb lattice (Fig.\ref{butt}) and in the previous case with a defect at $E$ (Fig.\ref{E}). In Hofstadter's original case on a square lattice, those points also correspond to the center of butterflies and their basic shapes are protected (compare also with Fig.\ref{2}) unless the fractality is lost. Therefore it would be natural to guess that such a gapless point exist without depending on lattice forms. This fact is very particular since the other gapless points which exist in the perfect lattice case disappear when there is a defect in a unit cell.

\section{\label{iii} Conclusion and Future Work}
In this note, we considered the Hofstadter problem on the honeycomb lattice with defects using the tight-binding Hamiltonian including the nearest-neighbor interactions. We examined the contributions from atoms to the formulation of Hofstadter's butterfly. Naively, increasing the number of defects in a unit cell make graphs less fractal. However we found a case where it recovers its fractal nature and acquires a new translation symmetry. In addition, we also surveyed the dependence on the size of unit cells with a single defect and unexpectedly found robust gapless points in the $E=0$ eV line. Those points exist without depending on the size of a unit cell and they exist at the center of butterflies. Therefore we may conclude that the butterfly at $\phi=0$ point is immortal as long as graphs have fractal structures.  

For future work, it will be interesting to consider Hamiltonian with the next-nearest-interactions. We found that when a unit cell has more than two defects, choices of missing points are crucial to energy structures. But we also admit that interactions among Bloch wave functions are also crucial. It would be also intriguing to find a topological method to describe butterflies with or without defects. Especially we are curious how the robustness of the gapless points can be described in a more theoretical way. When a standard topological insulator has a gapless point, then its topological number usually change there. We expect there would be a certain topological number whose parameter is $\phi$. Moreover, extension to higher dimensional case remains open question. \if{As far as authors' knowledge, the Hofstadter problem in higher dimensional spaces with defects has not studied yet.}\fi Does such a robust gapless point generally exist in higher dimensional lattice spaces? If so, does it depend on space's dimension? How can it be generically described in a uniformed manner?   

\acknowledgements
We wish to thank Mikito Koshino for useful discussions and advices. 
\bibliography{Ref}
\end{document}